\documentclass{iopart}

\usepackage{iopams}
\usepackage{hyperref}
\usepackage{graphicx}
\usepackage{mathptmx}
\usepackage[ngerman,american]{babel}
\usepackage{cite}
\usepackage{url}

\usepackage[dvipsnames,svgnames]{pstricks}
\usepackage{pstricks-add}
\usepackage{pst-3d}
\usepackage{pst-3dplot}
\usepackage{pst-all,pst-plot,pst-math,pst-poly,pst-node,pst-blur}

\renewcommand{\NP}{{\bf NP}} 
\newcommand{\PP}{{\bf P}}  
\renewcommand{\NC}{{\bf NC}}  
\newcommand{\p}{\tau} 
\newcommand{\pg}{p} 
\newcommand{\R}{S} 
\newcommand{\LL}{L} 
\newcommand{\ft}{\tau} 
\newcommand{\T}{D} 
\newcommand{\B}{{\mathcal B}} 
\newcommand{\G}{{\mathcal N}} 
\newcommand{\et}{T} 
\newcommand{\Ft}{P} 
\newcommand{\true}{T} 
\newcommand{\dx}{\mathrm{d}x}
\renewcommand{\e}{\mathrm{e}}
\newcommand{\bigo}{\mathcal{O}}
\newcommand{\TRUE}{TRUE}
\newcommand{\FALSE}{FALSE}
\newcommand{\f}{F}  
\newcommand{\rmax}{r_{\mathrm{max}}}
\newcommand{\poly}{\mathrm{poly}}
\newcommand{\eqref}[1]{(\ref{#1})}

\newcommand{\fuenf}[1]{#1 5 exp 1 #1 sub 5 exp add}

\savedata{\circuitAdata}[
4.85203 77.416
5.54518 93.062
6.23832 108.302
6.93147 122.52
7.62462 136.987
8.31777 155.394
9.01091 173.09
9.70406 189.94
]

\savedata{\circuitBdata}[
4.85203 57.227
5.54518 66.628
6.23832 77.527
6.93147 86.341
7.62462 98.153
8.31777 110.515
9.01091 120.845
9.70406 132.21
]

\savedata{\circuitCdata}[
4.85203 38.17
5.54518 43.734
6.23832 50.945
6.93147 55.842
7.62462 63.23
8.31777 69.541
9.01091 75.685
9.70406 82.38
]

\savedata{\circuitDdata}[
4.85203 21
5.54518 23.473
6.23832 26.546
6.93147 29.764
7.62462 33.286
8.31777 35.972
9.01091 39.351
9.70406 42.028
]

\savedata{\circuitmdata}[
0.505  1.97933
0.51  1.66655
0.52  1.36314
0.53  1.19237
0.54  1.0721
0.55  0.961136
0.56  0.886186
0.57  0.809977
0.6  0.647168
0.65  0.441408
0.7  0.284205
0.75  0.155661
0.8  0.0319253
]

\savedata{\critmdata}[
0.95 0.391415
0.9 0.659714
0.85 0.835754
0.8 0.998822
0.75 1.12724
0.7 1.20549
0.6 1.35142
0.5 1.40455
]

\savedata{\critAdata}[
100 39.081
200 82.652
300 110.402
400 150.847
500 193.725
600 251.37
700 292.2
800 297.039
900 347.057
1000 393.853
1100 433.644
1200 456.381
1300 439.309
1400 557.301
1500 603.702
1600 595.98
1700 665.831
1800 680.643
1900 757.895
2000 818.03
]

\savedata{\critBdata}[
100 62.632
200 118.468
300 187.267
400 256.142
500 335.019
600 377.151
700 443.292
800 535.698
900 582.141
1000 669.098
1100 753.105
1200 797.937
1300 816.999
1400 895.635
1500 976.87
1600 1091
1700 1184.32
1800 1178.01
1900 1217.71
2000 1278.94
]

\savedata{\critCdata}[
100 82.337
200 171.311
300 255.666
400 340.593
500 415.314
600 524.127
700 600.381
800 704.4
900 713.568
1000 792.49
1100 923.52
1200 988.301
1300 1075.69
1400 1174.79
1500 1357.49
1600 1330.98
1700 1431.61
1800 1468.89
1900 1601.04
2000 1664.61
]

\savedata{\critDdata}[
100 91.236
200 198.256
300 288.575
400 404.546
500 523.367
600 578.911
700 711.95
800 796.214
900 871.128
1000 939.313
1100 1113.62
1200 1240.01
1300 1302.66
1400 1391.98
1500 1509.15
1600 1573.75
1700 1627.69
1800 1785.52
1900 1875.64
2000 2063.5
]

\savedata{\critEdata}[
100 112.539
200 216.848
300 323.307
400 456.333
500 549.027
600 641.779
700 762.262
800 890.791
900 1031.19
1000 1115.68
1100 1252.79
1200 1353.55
1300 1496.95
1400 1568.99
1500 1740.09
1600 1754.99
1700 1922.29
1800 1984.99
1900 2121.66
2000 2250.76
]

\savedata{\critFdata}[
100 117.988
200 238.945
300 341.499
400 482.796
500 640.665
600 728.242
700 869.996
800 974.267
900 1088.42
1000 1340.72
1100 1340.02
1200 1479.84
1300 1599.62
1400 1707.66
1500 1771.52
1600 1931.64
1700 2111.82
1800 2180.41
1900 2258.64
2000 2406.41
]

\savedata{\critGdata}[
100 136.85
200 269.487
300 402.602
400 542.697
500 696.524
600 800.899
700 946.768
800 1059.67
900 1233.52
1000 1345.78
1100 1438.61
1200 1633.38
1300 1800.26
1400 1882.65
1500 2019.94
1600 2159.81
1700 2334.57
1800 2523.42
1900 2489.31
2000 2677.96
]

\savedata{\critHdata}[
100 128.474
200 269.222
300 406.748
400 550.623
500 699.139
600 847.549
700 972.715
800 1099.24
900 1230.13
1000 1400.39
1100 1529.52
1200 1664.64
1300 1836.07
1400 1913.38
1500 2105.32
1600 2169.16
1700 2422.68
1800 2510.18
1900 2653.2
2000 2823.09
]

\begin{document}

\title{Parallel Complexity of Random Boolean Circuits}

\author{ J Machta$^{1,2}$, S DeDeo$^1$, S Mertens$^{1,3}$ and C Moore$^{1,4}$}


\address{$^1$
Santa Fe Institute,
1399 Hyde Park Rd,
Santa Fe, NM 87501,
USA}

\address{$^2$
Physics Department,
University of Massachusetts,
Amherst, MA 010003, USA}

\address{
{\selectlanguage{ngerman}$^3$ Institut f\"ur Theoretische Physik,
Otto-von-Guericke Universit\"at,
D-39106 Magdeburg,
Germany}}

\address{$^4$
Computer Science Department and Department of Physics and Astronomy,
University of New Mexico,
Albuquerque, NM 87131, USA}

\begin{abstract}
Random instances of feedforward Boolean circuits are studied both analytically and numerically.  Evaluating these circuits is known to be a \PP-complete problem and thus, in the worst case, believed to be impossible to perform, even given a massively parallel computer, in time much less than the depth of the circuit. Nonetheless, it is found that for some ensembles of random circuits, saturation to a fixed truth value occurs rapidly so that evaluation of the circuit can be accomplished in much less parallel time than the depth of the circuit.  For other ensembles saturation does not occur and circuit evaluation is apparently hard.  In particular, for some random circuits composed of connectives with five or more inputs, the number of true outputs at each level is a chaotic sequence.  Finally, while the average case complexity depends on the choice of ensemble, it is shown that for \emph{all} ensembles it is possible to simultaneously construct a typical circuit together with its solution in polylogarithmic parallel time. 
\end{abstract}



\section{Introduction}
\label{sec:intro}

One of the most fruitful areas of interaction between statistical physics and computer science is the study of 
phase transitions in 
random instances of computational problems.  Using ideas from statistical physics, a number of studies have demonstrated that for many \NP-hard optimization problems, the running time of algorithms for solving these problems is largest at a phase transition separating a regime where a solution exists from a regime where no solution exists~\cite{MoZeKiSeTr99,mezard:montanari,moore:mertens}.  

It has also been found that for some random ensembles of hard problems, if the problem is sufficiently unconstrained, then we can sample from the distribution of (instance, solution) pairs, at least approximately, using the so-called \emph{planted ensemble}~\cite{achlioptas:coja-oghlan:08,KrZd09}.  That is, rather than choosing a random problem and then solving it, we first choose a random solution, and then choose randomly from among the instances consistent with it.  This approach is especially relevant to statistical physics, since we are more interested in averaging over ensembles of disordered systems than in solving individual hard problems.

While considerable effort has been expended on understanding random ensembles of  \NP-hard problems, we are unaware of comparable investigations lower in the complexity hierarchy.  In this paper we investigate both phase transitions in complexity and sampling (instance, solution) pairs for random ensembles of the circuit value problem, a problem in the class \PP.

Contained within the class \PP\ of problems that are solvable in
polynomial time there are several nested hierarchies of complexity
classes that are best understood in terms of parallel
computation~\cite{GrHoRu}.  For our purposes, the most important of
these classes is \NC, the class of problems that can be solved
\emph{in parallel} in polylogarithmic time, i.e., $\bigo(\log^k N)$
time for some constant $k$, where $N$ denotes the size of the problem. In the definition of \NC\ the model of computation that is considered is the PRAM, an idealized parallel computer with a number of processors that is allowed to scale polynomially in the size of the problem and with a global random access memory through which any pair of processors can communicate in $\bigo(1)$ time.  The PRAM runs synchronously and each processor runs the same program but has a distinct label so that it may carry out distinct computations.  In each time step, each processor may read or write to a global memory cell.  Conflicts that arise if two processors attempt to write to the same memory cell at the same time may be resolved in different ways, but these differences do not change the definition of the class \NC. Although we have just defined \NC\ in terms of a specific model of parallel computation, it is a quite robust class of problems and can be equivalently defined in terms of families of Boolean circuits, alternating Turing machines and even, without reference to computation at all, in terms of properties of the first order formal logic description of the problem~\cite{Im99}.

Problems in \NC\ include adding $N$ numbers, multiplying two $N \times N$ matrices, and finding the connected components of a graph with $N$ vertices.  In the case of addition, the parallel algorithm consists simply of pairwise addition of $N$ numbers by $N/2$ processors, followed by pairwise addition of these partial sums by $N/4$ processors, and so on until the sum is obtained after $\log_2 N$ parallel steps.  

A question then arises as to whether every problem in \PP\ can be solved in parallel in polylogarithmic time---that is, whether $\PP=\NC$.  It is widely believed but not yet proved that this is not the case, and that there are problems in \PP\ that are hard to solve in parallel.  Just as \NP-complete problems are the hardest problems in the class \NP, there is a class of \PP-complete problems that are the hardest problems in \PP\ to solve in parallel.  Assuming that $\PP\neq\NC$, \PP-complete problems are \emph{inherently sequential}---they cannot be solved in polylogarithmic time on a PRAM with polynomially many processors.  

The canonical \PP-complete problem is the Circuit Value Problem (CVP).  An instance of CVP is specified by a feedforward Boolean circuit with given truth values for the inputs, and the problem is to find the outputs.  If the circuit has $N$ gates, then this problem is clearly in \PP\ since we can evaluate the output of each gate in roughly $\bigo(N)$ time.  The question is to what extent we can parallelize this computation.  The \emph{depth} of a circuit is the longest path from an input to an output.  By evaluating all the gates in a given layer simultaneously, then we can solve CVP in an amount of parallel time proportional to the depth.  However, it is not clear that we can improve significantly on this, and since a circuit with $N$ gates could have, say, $\sqrt{N}$ depth and $\sqrt{N}$ width, it seems unlikely that we can solve CVP in $\bigo(\log^k N)$ parallel time.  Any such algorithm would have to ``skip over'' many of the layers of a circuit, rather like predicting the future state of a system without having to simulate it step-by-step.

CVP plays the same central role in the theory of \PP-completeness as Satisfiability plays in the theory of \NP-completeness.  Reductions from CVP or Satisfiability are the standard tools for proving other problems \PP-complete or \NP-complete, respectively.  Just as a polynomial-time algorithm for Satisfiability would imply that $\PP=\NP$, if CVP is in \NC\ then $\PP=\NC$ and all problems that can be solved in polynomial time can be efficiently parallelized.

In this paper we demonstrate the existence of hard and easy phases and
transitions between them in random ensembles of CVP.   

We show that random monotone CVP is easy if the circuit almost surely saturates quickly to one of the two stable fixed points--either all \TRUE\ or all \FALSE.  Random NOR CVP quickly saturates to a period-two oscillation between \TRUE\ and \FALSE.  Our analysis uses simple, exact recursion relations for the expected number of \TRUE\ outputs at one level as a function of the fraction at the previous level.  Similar methods were employed by Valiant~\cite{Va84} to demonstrate the existence of monotone circuits that quickly evaluate the majority function.   In addition to circuits composed of simple gates (i.e., AND, OR, NOR, and NAND gates) we also consider random circuits built out of more complicated Boolean functions or \emph{connectives}.  Here we find examples where the fraction of \TRUE\ outputs of each level of the circuit is a chaotic sequence; this leads to a qualitatively different way that CVP can become hard.  Finally, we show that for \emph{any} choice of parameters it is possible to simultaneously construct problem instances and their solutions quickly in parallel.  It is thus easier to sample (instance, solution) pairs than to be given an instance and then solve it.

Random Boolean circuits have a long history.  Early work by von
Neumann~\cite{vN56} and Moore and Shannon~\cite{MoSh56} focused on
circuits with unreliable gates.  Recursion relation methods were
developed in~\cite{MoSh56} that parallel the methods used here.  Later
work by Valiant~\cite{Va84} used random monotone circuits to compute
the majority function.  This work was extended to study other
functions that can be computed by random circuits generated via a
growth process, see for example~\cite{BrPi05} and references therein.
In~\cite{BrPi05} the recursion relation is called a characteristic
polynomial and the elementary Boolean function is referred to as the
`connective,' a term we reserve here for more complicated Boolean
functions composed of multiple AND, OR, NAND, or NOR gates. Recent
work by~\cite{Moz10} considers the joint problem of random circuits
and unreliable gates, finding associations between reliability bounds
and macroscopic phase transitions.  In our ensembles, each layer of
the circuit is chosen simultaneously, and the circuit has fixed width;
for another ensemble of random circuits, where random gates are added
one at a time, see~\cite{diaz:etal}.

The paper is organized as follows. In Sec.~\ref{sec:cvp} we introduce random ensembles of Boolean circuits.  In Sec.~\ref{sec:monotone} we analyze the difficulty of solving monotone CVP as a function of the in-degree of the gates and other parameters describing the ensemble and demonstrate the existence of different phases of hardness.  Section~\ref{sec:nor} considers the case of NOR CVP.   In Section~\ref{sec:chaos} we show how chaotic behavior in circuit properties is possible for connectives of sufficiently high in-degree. Section~\ref{sec:num} presents numerical results supporting the theoretical conclusions for monotone circuits.  In Sec.~\ref{sec:sampling} we show that (instance, solution) pairs can be sampled in polylogarithmic parallel time even in the phases where solving giving random instances is hard.  The paper concludes in Sec.~\ref{sec:disc} with a summary and discussion.

\section{Random Circuit Value Problems}
\label{sec:cvp}

We define random Boolean circuits as follows.  We think of them as lying on a rectangular grid of width $\LL$, although the circuit's topology is less restricted than the grid would suggest.  On the top row, the $\LL$ inputs of the circuit are independently chosen and take the value \TRUE\ with probability $\p_0$ and \FALSE\ with probability $1-\p_0$.  There are $\LL$ gates (or connectives) on each level, each of which has some in-degree $k$.  Each gate on level $n$ takes inputs from $k$ randomly chosen gates on level $n-1$, and a gate on the first level takes inputs from $k$ randomly chosen input values.  Note that the inputs to a gate are chosen randomly with replacement so that inputs may be repeated.  

\begin{figure}
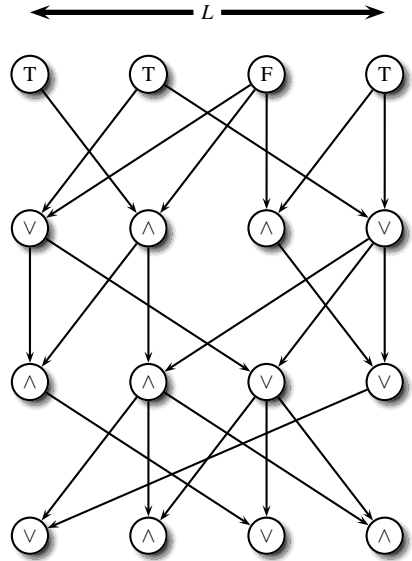

\begin{indented}
  \item[] \vspace{8mm}
  \begin{center}
   \psmatrix[mnode=circle,colsep=0.1\linewidth,shadow=true,blur=true,shadowsize=3pt,blurradius=2pt]
  [mnode=r]  & & & [mnode=r] \\[-8ex]
  T & T & F & T \\
  $\vee$ & $\wedge$ & $\wedge$ & $\vee$ \\
  $\wedge$ & $\wedge$ & $\vee$ & $\vee$ \\
  $\vee$ & $\wedge$ & $\vee$ & $\wedge$ 
  \endpsmatrix
  \psset{shadow=false,arrows=<-}
  \ncline[arrows=<->,linewidth=2pt]{1,1}{1,4}\ncput*{$\LL$}
  \ncline{3,1}{2,2} \ncline{3,1}{2,3} \ncline{3,2}{2,1} \ncline{3,2}{2,3} \ncline{3,3}{2,3} \ncline{3,3}{2,4} \ncline{3,4}{2,2} \ncline{3,4}{2,4} 
  \ncline{4,1}{3,1} \ncline{4,1}{3,2} \ncline{4,2}{3,2} \ncline{4,2}{3,4} \ncline{4,3}{3,1} \ncline{4,3}{3,4} \ncline{4,4}{3,3} \ncline{4,4}{3,4} 
  \ncline{5,1}{4,2} \ncline{5,1}{4,4} \ncline{5,2}{4,3} \ncline{5,2}{4,2} \ncline{5,3}{4,3} \ncline{5,3}{4,1} \ncline{5,4}{4,2} \ncline{5,4}{4,3} 
  \end{center}
\end{indented}
\caption{An example of a random monotone Boolean circuit arranged in levels on a rectangular grid of width $\LL$ with each gate having $k=2$ inputs.  Each of the $k$ inputs of a gate is attached to a randomly chosen gate at the next higher level of the circuit.}
\label{fig:rancvp}
\end{figure}

We consider monotone circuits, circuits consisting entirely of NOR gates, and circuits built from more complicated connectives.  We start with monotone circuits, which contain only AND ($\land$) and OR ($\lor$) gates.   Each gate is OR with probability $\pg$, or AND with probability $(1-\pg)$.  These gates can have any in-degree $k$; OR gates return \TRUE\ if at least one of their inputs are \TRUE, and AND gates return \TRUE\ if all of their inputs are \TRUE.  We will explicitly study two-input and three-input gates, i.e., $k=2$ and $k=3$; the case case $k>3$ is very similar to $k=3$. A typical monotone circuit with two-input gates is shown in Figure~\ref{fig:rancvp}.

Our goal is to understand the parallel complexity of the CVP problem in this ensemble, i.e., the complexity of computing the truth values on the final layer as a function of the inputs.  This problem is \PP-complete whenever $\LL$ is allowed to grow as some power of the total number of gates.  Even Monotone CVP is \PP-complete~\cite{GrHoRu}, since we can use de Morgan's law to push negations up through the circuit, making the gates monotone and flipping some of the gates.  Since $\mathrm{NOR}(x,y)=\lnot (x \lor y)$ is a complete basis for Boolean logic, i.e., any Boolean function can be implemented using only NOR gates, NOR CVP is also \PP-complete.

\section{Analysis of Random Monotone CVP}
\label{sec:monotone}

In monotone circuits, the homogenuous states (all \TRUE\ or all
\FALSE) are absorbing states in the sense that if they appear in one
layer of the circuit, they will persist all the way to the final
layer, independently of the wiring and the gates. Therefore the parallel
computational complexity is determined by the time to reach one of
these absorbing states.

Let $\ft_n$ denote the fraction of the gates at level $n$ whose output is \TRUE.  Because of the random connectivity of the circuit, it is possible to write the expectation of $\ft_{n+1}$ exactly as a function of $\ft_n$.  For the case of monotone circuits with two inputs, $k=2$, this is
\begin{eqnarray}
\label{eq:rec2}
\langle \ft_{n+1} \rangle &=& \pg(2\ft_n-\ft_n^2) + (1-\pg) \ft_n^2 
\\
\nonumber
&=& \ft_n + (2 \pg-1)( \ft_n-\ft_n^2) \, .
\end{eqnarray}
The first term in the first expression is the probability that an OR gate will evaluate to \TRUE\ times the expected fraction $\pg$ 
of OR gates, and the second term is the analogous quantity for AND gates.  More generally, for $k$-input AND and OR gates we have
\begin{equation}
\label{eq:reck0}
\langle \ft_{n+1} \rangle = R_k(\ft_n,\pg) \equiv \pg \big( 1-(1-\ft_n)^k \big) + (1-\pg)\ft_n^k \, .
\end{equation}
In the limit of large $L$, $\ft_{n+1}$ is tightly concentrated, 
allowing us to drop the distinction between the random variable
$\ft_{n+1}$  and its expectation $\langle\ft_{n+1}\rangle$. This gives a recursion relation
\begin{equation}
\label{eq:reck}
\ft_{n+1} = R_k(\ft_n,\pg) \, .
\end{equation}

The initial condition for this recursion relation is the fraction of \TRUE\ inputs, $\p_0$.  The endpoints $\ft=0$ and $\ft=1$ are fixed points of the recursion relations.  Once the circuit saturates at $\ft=0$ or $\ft=1$, any further levels simply reproduce these truth values.  Thus the depth to reach saturation is an upper bound on the parallel time required to evaluate the circuit.  As we see below, this saturation depth depends on the fluctuations around $\langle \ft_{n+1} \rangle$ for finite $L$.

\subsection{Two-input gates}
\label{sec:2in}

Consider the case $k=2$.  The only fixed points are $\ft=0$ and $\ft=1$.  From the second equality in Eq.~\eqref{eq:rec2}, it is clear that the behavior of the recursion relations changes as a function of $\pg$ at $\pg=1/2$.  For $\pg<1/2$, $\ft=0$ is stable and $\ft=1$ is unstable, and the flow is from $\ft=1$ to $\ft=0$.  The opposite holds for $\pg>1/2$. 

We can draw the following qualitative conclusions from these flows.  A sufficiently deep random circuit with a preponderance of AND gates will almost always evaluate to \FALSE, while one with a preponderance of OR gates will almost always evaluate to \TRUE.  Let us define the saturation depth $\T$ as the mean level $n$ at which the circuit first saturates (almost always \TRUE\ for $\pg>1/2$ and \FALSE\ for $\pg<1/2$).  We can estimate $\T$ as the least $n$ for which $\ft_n \approx 1/\LL$.  

First consider the case $\pg<1/2$.  Linearizing around the relevant fixed point at $\ft=0$, we obtain $\ft_{n+1} = 2\pg \ft_n$ and so $\ft_{n} = (2\pg)^n \p_0$.  This yields the estimate
\begin{equation}
\label{eq:plh}
  \T \sim \frac{\ln \LL}{-\ln(2\pg)} \, .
\end{equation}
A similar calculation for $\pg>1/2$, obtained by linearizing around the fixed point at  $\ft=1$, yields
\begin{equation}
\label{eq:pgh}
 \T \sim \frac{\ln \LL}{-\ln(2(1-\pg))}.
\end{equation}
In either case, $\T$ diverges like $1/|\pg-1/2|$ as $\pg \rightarrow 1/2$.

The case $\pg=1/2$ cannot be understood in terms of the recursion relation~\eqref{eq:rec2} for the expected fraction of \TRUE\ gates.  Instead of 
recursion relations for the expectation $\ft_n$,
we need to follow the stochastic behavior of $\et_n$, the actual number of \TRUE\ gates at level $n$. Including fluctuations in the fraction of gates of each type on a given level, the distribution of $\et_{n+1}$ is a single binomial distribution whose mean is obtained from $\et_n$ using the recursion relation, Eq.~\eqref{eq:rec2} with $\et_n/L$ replacing $\ft_n$.  In the special case that $\pg=1/2$, the recursion relation is the identity and
\begin{equation}
\label{eq:bin}
\et_{n+1}=\B(\LL,\et_n/\LL) \, . 
\end{equation}
Here $\B(N,p)$ is a binomial random variable whose value is the number of successes in $N$ trials with probability of success $p$.  The initial condition for this recursion relation is $\et_0=\B(\LL,\p_0)$.  

Equation~\eqref{eq:bin} describes a random walk with variable step length on a line with absorbing states at 0 and $\LL$.  In order to analyze this walk we take the large $\LL$ limit and replace the binomial by a normal random variable,
\begin{equation}
\label{eq:norm}
\et_{n+1}=\G(\et_n,\et_n(1-\et_n/\LL))
\end{equation}
where $\G(\mu,\sigma^2)$ is a normal random variable with mean $\mu$ and variance $\sigma^2$.   Let  $\Ft(x,n)$ be the probability density for the number of \TRUE\ gates at level $n$, $\Ft(x,n) \,\dx={\rm Prob}\left[\et_n\in (x,x+\dx) \right]$. The recursion relation for $\Ft$ follows from Eq.~\eqref{eq:norm} and takes the form of an integral equation with a diffusion Green's function with a spatially-varying diffusion coefficient,
\[
\Ft(x,n+1)=\int_0^\LL \dx' \,G(x,x') \Ft(x',n) \, , 
\]
where $G$ is the one-dimensional diffusion kernel
\[
G(x,x') =\frac{1}{ \sqrt{4 \pi K(x')}} \,\e^{-(x-x')^2/4K(x')}
\]
and $K(x)$ is the diffusion coefficient
\begin{equation}
\label{eq:D}
K(x)= \frac{x}{2}\left(1-\frac{x}{\LL}\right).
\end{equation}

The saturation depth $\T$ at which the gates are all \TRUE\ or all \FALSE\ is the mean first-passage time to the absorbing states at $x=0$ or $x=L$.  The  first-passage time for a diffusion process with a diffusion coefficient that varies as in Eq.~\eqref{eq:D} is analyzed in~\cite{redner01}.  The result depends on the initial condition.  In our case, $\Ft(x,0)=\delta(x-\p_0\LL)$, where $\p_0$ is the expected fraction of \TRUE\ inputs.  After appropriate changes of variable to put our expression in the form given in Section 4.6.2 of~\cite{redner01}, we obtain~\footnote{Our results follows from Eq. (4.6.6) of~\cite{redner01} as corrected in the online errata.}
\begin{equation}
\label{eq:peh}
\T= -2\LL\left[\p_0\ln\p_0+(1-\p_0)\ln(1-\p_0)\right] 
= 2 \LL h(\p_0) \, , 
\end{equation}
where $h(\tau) = -\tau \ln \tau - (1-\tau) \ln (1-\tau)$ is the Gibbs-Shannon entropy of the initial distribution of inputs.  Note that $D$ is now linear in the width of the circuit instead of logarithmic, as was the case for $\pg \neq 1/2$, suggesting that the parallel complexity is $\bigo(L)$ rather than $\bigo(\log L)$.

\subsection{Three-input gates}

Next consider the case where each gate has 3 inputs.  The fixed point structure of the recursion relations for $k=3$ is more complicated than for $k=2$.  In addition to the fixed points at $\ft=0$ and 1, there is a third fixed point at $\ft^\ast=3\pg-1$.  
This fixed point is meaningful only for $1/3<\pg<2/3$, since otherwise it is outside the unit interval.  

For $\pg<1/3$, it is straightforward to verify that $\ft_{n+1} < \ft_n$ so that $\ft=0$ is the stable fixed point and $\ft=1$ the unstable fixed point.  Thus the regime $\pg<1/3$ for 3-input gates is similar to the regime $\pg<1/2$ for 2-input gates: in both cases the circuit almost always saturates to \FALSE\ at depth $\T =\bigo(\log L)$.   For $\pg>2/3$, $\ft=1$ is the stable fixed point and the circuit saturates to \TRUE, again at logarithmic depth.  

The regime $1/3<\pg<2/3$ has no analogy for 2-input gates.  Linearizing the recursion relation around the fixed point by setting $\ft_n=\ft^\ast+\delta \ft_n$ in Eq.~\eqref{eq:reck}, 
we obtain 
\begin{equation}
\label{ }
\delta \ft_{n+1} = (9 \pg^2 - 9 \pg +3) \delta \ft_n \, .
\end{equation}
The coefficient of $\delta \ft_n$ in this equation is between 0 and 1, 
implying that the fixed point $\ft^\ast=3\pg-1$ is stable while the fixed points at $\ft=0,1$ are unstable.  The conclusion is that for any $1/3<\pg<2/3$ and $0<\p_0<1$ the circuit fails to saturate, but instead behaves stochastically for a very long time, with $\ft$ making $\bigo(1/\sqrt{L})$ fluctuations around $\ft^\ast$.  
For any finite $\LL$ it is possible for the system to fall into one of the absorbing states, either all \TRUE\ or all \FALSE, but at each step this is exponentially unlikely.  Thus for exponentially long time scales, saturation will almost surely occur, but for depths and widths that are related polynomially, the circuit will simply have to be evaluated one level at a time.

\subsection{Gates with more than three inputs}

The case $k>3$ is qualitatively similar to the case $k=3$. For $\pg < 1/k$, the only stable fixed point is $\ft = 0$, and for $\pg > 1-1/k$ the only stable fixed point is $\ft = 1$.  In both cases, the circuit saturates at logarithmic depth.  

For $1/k < \pg <1-1/k$, on the other hand, both these fixed points become unstable, and there is a single attracting fixed point $0 < \ft^\ast < 1$.  The recursion $R_k(\ft,\pg)$ 
rises sharply from $\ft=0$, is relatively flat near $R_k(\ft,\pg) \approx \pg$, and then rises sharply again to one near $\ft=1$.  Thus there is one stable fixed point at $\ft^\ast\approx \pg$, and it becomes increasingly stable as $k$ increases.  The saturation depth is again exponential, and these cases of CVP are hard.

\section{Analysis of Random NOR CVP}
\label{sec:nor}

The defining feature of the monotone ensembles considered above is the absence of negation: increasing the fraction of \TRUE\ inputs increases the likelihood of \TRUE\ outputs.   
In this section we consider random ensembles with non-monotone gates, namely NOR CVP.  As mentioned above, NOR CVP is \PP-complete, since NOR is a complete basis for Boolean logic.

We define a random ensemble of NOR gates on a grid of width $L$.  Each gate takes $k$ inputs, chosen randomly with replacement, from the level above it.  Initially, there is a fraction $\p_0$ of \TRUE\ inputs.  Since a NOR gate returns \TRUE\ if and only if all its inputs are false, the recursion relation is
\begin{equation}
\label{eq:norreck}
\ft_{n+1}=\R_k(\ft_n)\equiv(1-\ft_n)^k .
\end{equation}

First, consider the case $k=2$.  The recursion relation has a single fixed point in the unit interval at $\ft^\ast=(3-\sqrt{5})/2\approx0.382$.  
Since $|d\R_2(\ft^\ast)/d\ft| = \sqrt{5}-1>1$, this  fixed point is unstable and flows to a stable period-two orbit, oscillating between $\ft=0$ and $\ft=1$.  
This orbit is stable, since $\R(\R(\ft)) = 4 \ft^2 + \bigo(\ft^3)< \ft$ for sufficiently small $\ft$.  

Once the circuit reaches this fixed point, it has saturated in a way analogous to the monotone ensemble of the previous section.  The gates alternate between all \TRUE\ and all \FALSE, and no additional computation is needed to predict the behavior deeper in the circuit.  Since the period-two orbit is approached exponentially, and saturation occurs when $\ft_n \approx 1/\LL$, the saturation depth is $\T \sim \log \LL$ just as in the easy regime of monotone CVP.  This is the case even if $\p_0$ starts at the unstable fixed point $\ft^\ast$, since random fluctuations cause $\ft$ to vary by $\bigo(1/\sqrt{\LL})$ in any case.  The expected distance from $\ft^\ast$ grows exponentially, driving $\ft$ a distance $\bigo(1)$ away from $\ft^\ast$ after $\bigo(\log \LL)$ steps.  Thus the entire phase diagram for NOR CVP is easy. 

Unlike monotone CVP, the case $k=2$ is generic for NOR CVP.  For any $k$ the period-two orbit is stable since $\R_k(\R_k(\ft))= (k \ft)^k + \bigo(\ft^{k+1})< \ft$ for $\ft$ sufficiently small.  Since $\R_k(x)$ is a convex function that varies between one at $x=0$ and zero at $x=1$, there is exactly one fixed point $\ft^\ast$ in the unit interval; it is unstable for all $k$, and its value tends toward zero as $k$ increases.  Apart from these quantitative differences, the  behavior of $k$-input random NOR circuits is the same for all $k$.  The saturation depth is logarithmic, and predicting the behavior of the circuit is easy for all $k$ and all $\p_0$.

\section{Chaotic Connectives}
\label{sec:chaos}

While the recursion relations Eqs.~\eqref{eq:reck} and~\eqref{eq:norreck} corresponding to simple AND, OR, and NOR gates have simple dynamics, with a handful of fixed points and stable periodic orbits, connectives that implement more complicated Boolean functions can produce a variety of  complicated behaviors.  In this section we show an example of such a connective whose dynamics are chaotic, requiring a step-by-step simulation even to keep track of the expected fraction of \TRUE\ outputs.

Since the circuit is wired randomly, many connectives -- those whose truth table entries can be related by permutations of input bits -- produce equivalent recursion relations.  For a connective of in-degree $k$, the full space of recursion relations is thus covered by $k+1$ integer parameters, $0 \le \alpha_i \le {k \choose i}$, which define how many input bit configurations with $i$ \TRUE\ bits lead to a \TRUE\ output.  The recursion relation for such a connective is
\begin{equation}
\label{eq:karyreck}
\ft_{n+1} = \sum_{i=0}^k \alpha_i \ft_n^{i}(1-\ft_n)^{k-i} \equiv \f(\ft_n) \, .
\end{equation}
If the layers consist of a mixture of different connectives with the same in-degree $k$, as in the mix of AND and OR gates studied in Sec.~\ref{sec:monotone}, the $\alpha_i$ become weighted averages and can take arbitrary real values between $0$ and ${k \choose i}$.

Varying the $\alpha_i$ lets us construct connectives, or mixtures of connectives, such that the recursion Eq.~\eqref{eq:karyreck} becomes a wide variety of functions $\f(\ft)$ on the unit interval.  For instance, consider the logistic map,
\begin{equation}
\label{logistic}
f(\ft) = r \ft (1-\ft) \, .
\end{equation}
This undergoes a series of period-doubling transitions as $r$ increases, and becomes chaotic at $r \approx 3.57$ (see, e.g., Ref.~\cite{Strogatz94}).  Setting Eq.~\eqref{eq:karyreck} equal to Eq.~\eqref{logistic} and solving for the $\alpha_i$, we find the logistic map can be reproduced by taking
\begin{equation}
\label{kprecise}
\alpha_i = \left\{ \begin{array}{cl}
 0 &\mbox{if $i=0$ or $i=k$,} \\
 r{k-2\choose i-1} &\mbox{if $0< i< k$.} 
\end{array}\right.
\end{equation}
Connectives of this form are non-monotone and XOR-like, peaking at $\ft=1/2$.  Note that there is an absorbing state, i.e., the fixed point $f(0)=0$.

Because of the upper bound on $\alpha_i$, the maximum allowable value $\rmax$ of the parameter $r$ is a function of the size of the connective.  For $k=2$ we have $\rmax=2$, which corresponds to the standard XOR function.  More generally, we have
\begin{equation}
\label{maxr}
\rmax = \frac{2(2\lceil\frac{k}{2}\rceil-1)}{\lceil\frac{k}{2}\rceil} \, ,
\end{equation}
where $\lceil x \rceil$ is the smallest integer greater than or equal to $x$.  As $k$ increases, $\rmax$ increases and approaches $4$, giving us the full range of logistic maps.  In particular, we can reach the chaotic range $r \gtrsim 3.57$ with connectives of in-degree $k \ge 9$.

However, the binomials ${k-2 \choose i-1}$ appearing in~\eqref{kprecise} have no common factors. Therefore, for an arbitrary $r$ -- and, in particular, for $r$ in the chaotic regime $r\gtrsim3.5$ -- the $\alpha_i$ cannot be integers, and reproducing the logistic equation exactly requires a mixture of at least two connectives of sufficient $k$.

Given this restriction, one might ask whether it is possible to achieve chaotic behavior similar to that of the logistic map with a single connective.  One approach is to approximate the logistic map by rounding the $\alpha_i$ to nearby integers.  Given the universality of the period-doubling route to chaos, if the resulting approximation is good enough, the behavior will be similar.  For $k=9$ and $r = \rmax = 18/5$, we can take
\begin{equation}
\label{kapprox}
\alpha_i = \left\{ \begin{array}{rl}
 0 &\mbox{ if $i=0$ or $i=9$,} \\
 \lceil\frac{18}{5}{7\choose i-1}\rceil &\mbox{  $0< i< 9$.}
 \end{array} \right.
\end{equation}
and iterating the map for a range of different initial conditions shows chaotic behavior. We can measure the Lyapunov exponent $\lambda = \langle\log{|d\R/d\tau|}\rangle$, where averages are taken over the courses of long trajectories, and find that $\lambda \approx 0.14 > 0$. For comparison, the logistic map with $r=18/5$ has $\lambda \approx 0.18$.

Instead of approximating the logistic map, we can look directly for chaotic behavior via an exhaustive search of all possible connectives.  Since connectives whose truth tables are related by a permutation of the input are equivalent, there are far fewer than $2^{2^k}$ connectives to check; the number of distinct connectives is
\begin{equation}
n(k)=\prod_{i=0}^k \left[{k\choose i} + 1\right].
\end{equation}
Searching the maps associated with each connective for those with positive Lyapunov exponents, we find the first chaotic maps at $k=5$.   
Of the $17424$ distinct connectives for $k=5$, there are six chaotic ones.  The simplest one consists of a NOR gate and an AND gate, combined with an OR gate:
\[
\alpha_i = \left\{\begin{array}{cl}
 1 &\mbox{if $i=0$ or $i=5$,} \\
 0 &\mbox{if $0<i<5$.} 
       \end{array}\right. 
\]
This gives the recursion relation 
\begin{equation}
\label{kapprox_5}
\f(\ft) = \ft^5 + (1-\ft)^5 \, ,
\end{equation}
which has a Lyapunov exponent $\lambda \approx 0.20$.  
It is a unimodal map in the chaotic regime, with unstable orbits of period $1$, $2$, $4$, and so on.  Its first, second and fourth iterates are shown in Fig.~\ref{fivegate_iterated}.  

This map has an absorbing state at $\f(1) = 1$, which it can fall into due to finite-size fluctuations.  But since its chaotic attractor is bounded away from $1$, in the interval $[\f(1/2), \f(\f(1/2))] = [0.0625,0.7242]$, it will take exponential time for this to occur.  Thus the saturation depth $\T$ is exponential in $\LL$, and the CVP requires step-by-step simulation.

\begin{figure}
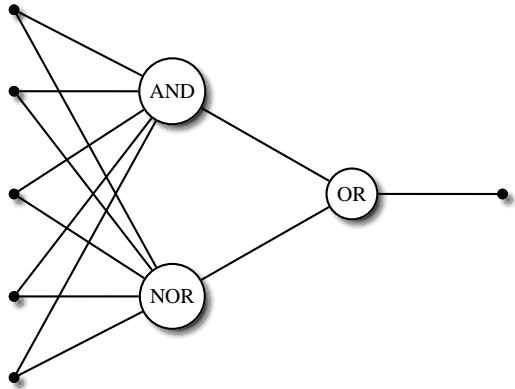

\begin{indented}
  \item[] \vspace{8mm}
  \begin{center}
   \psmatrix[mnode=circle,colsep=0.15\linewidth,rowsep=0.05\linewidth,shadow=true,blur=true,shadowsize=3pt,blurradius=2pt]
  [mnode=C,fillcolor=black,fillstyle=solid,radius=2pt] &  & & [mnode=r] \\
  [mnode=C,fillcolor=black,fillstyle=solid,radius=2pt] & AND &   &   [mnode=r] \\
  [mnode=C,fillcolor=black,fillstyle=solid,radius=2pt] &  &  OR & [mnode=C,fillcolor=black,fillstyle=solid,radius=2pt] \\
  [mnode=C,fillcolor=black,fillstyle=solid,radius=2pt] & NOR &   & [mnode=r] \\
  [mnode=C,fillcolor=black,fillstyle=solid,radius=2pt] &  &   & 
  \endpsmatrix
  \psset{shadow=false}
  \ncline{2,2}{3,3} \ncline{4,2}{3,3} \ncline{3,3}{3,4} 
  \ncline{1,1}{2,2} \ncline{2,1}{2,2} \ncline{3,1}{2,2} \ncline{4,1}{2,2} \ncline{5,1}{2,2} 
  \ncline{1,1}{4,2} \ncline{2,1}{4,2} \ncline{3,1}{4,2} \ncline{4,1}{4,2} \ncline{5,1}{4,2} 
  \end{center}
\end{indented}
\caption{The logic circuit diagram corresponding to Eq.~\eqref{kapprox_5}, one of the minimal arity connectives that produces chaotic behavior.}
\label{fivegate_picture}
\end{figure}

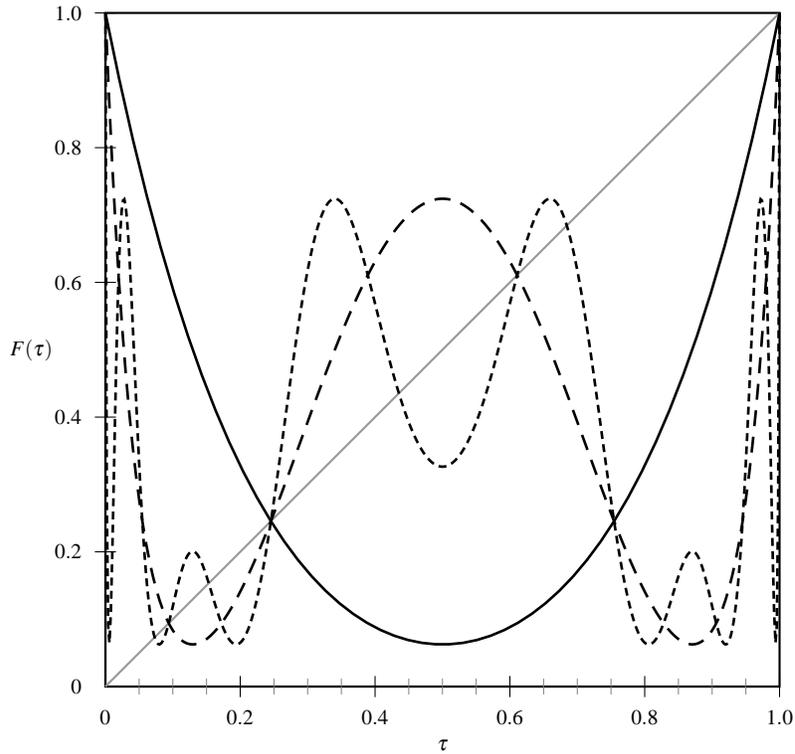
\begin{figure}
\begin{indented}
     \item[]
     \psset{unit=0.85\linewidth}
     \begin{pspicture}(-0.13,-0.08)(1.01,1.05)
       \psaxes[xsubticks=4,Dy=0.2,Dx=0.2,axesstyle=frame,showorigin=true]{->}(1.0,1.0)
       \uput{20pt}[-90](0.5,0){$\ft$}
       \uput{20pt}[180](0,0.5){$F(\ft)$}
       \psline[linecolor=black!40](1,1)
       \psset{linewidth=1pt}
       \psplot{0}{1}{\fuenf{x}}
       \psplot[linestyle=dashed,dash=6pt 4pt,plotpoints=200]{0}{1}{\fuenf{\fuenf{x}}}
       \psplot[linestyle=dashed,dash=3pt 2pt,plotpoints=500]{0}{1}{\fuenf{\fuenf{\fuenf{\fuenf{x}}}}}
     \end{pspicture}
   \end{indented}
\caption{Iterated maps of the chaotic connective, Eq.~\eqref{kapprox_5} and Fig.~\ref{fivegate_picture}, showing unstable periodic orbits of period $1$ (solid line), $2$ (long dashed line), and $4$ (short dashed line.)}
\label{fivegate_iterated}
\end{figure}

It is also possible to have combinations of connectives such that there is no absorbing state---that is, such that neither $\ft=0$ and $\ft=1$ is a fixed point, nor do they form an orbit of period two.  In that case the circuit never saturates, and the CVP is again hard.

\section{Numerical Results for Two Input Monotone CVP}
\label{sec:num}

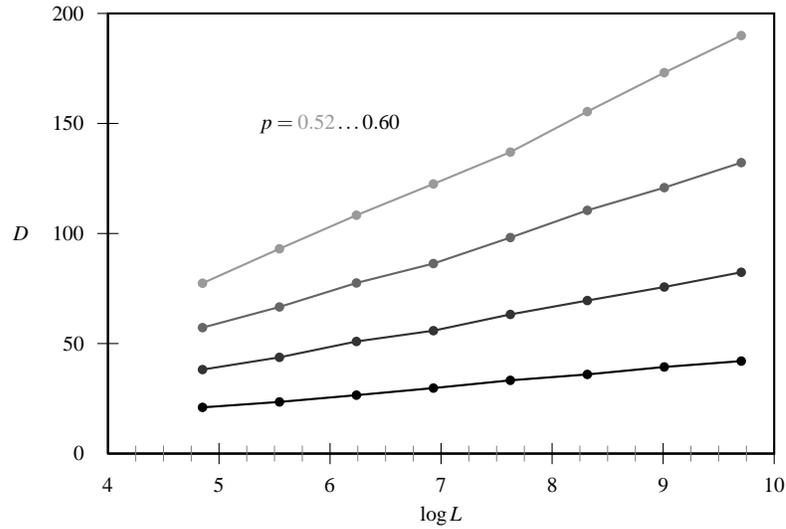
\begin{figure}
\begin{indented}
     \item[]
     \psset{xunit=0.14\linewidth,yunit=0.0027811\linewidth}
     \begin{pspicture}(3,-25)(10,200)
       \psaxes[xsubticks=4,Dy=50,Dx=1,axesstyle=frame,showorigin=true,Ox=4]{->}(4,0)(10,200)
       \uput{20pt}[-90](7,0){$\log\LL$} 
       \uput{30pt}[180](4,100){$\T$}
       \rput(6,150){$\pg={\color{black!40}0.52}\ldots 0.60$}
       \listplot[showpoints,linecolor=black!40]{\circuitAdata} 
       \listplot[showpoints,linecolor=black!60]{\circuitBdata} 
       \listplot[showpoints,linecolor=black!80]{\circuitCdata} 
       \listplot[showpoints,linecolor=black]{\circuitDdata} 
    \end{pspicture}
   \end{indented}
\caption{Saturation depth $\T$ vs. the log of the circuit width $\LL$
  for $k=2$ and various values of the fraction of OR gates $\pg$.}
\label{fig:circuit}
\end{figure}

\begin{figure}
\begin{indented}
     \item[]
     \psset{xunit=2.85\linewidth,yunit=0.26\linewidth}
     \begin{pspicture}(0.46,-0.3)(0.8,2)
      \psaxes[ylogBase=10,xsubticks=4,Dy=1,Dx=0.1,axesstyle=frame,showorigin=true,Ox=0.5]{->}(0.5,0)(0.801,2)
       \uput{20pt}[-90](0.65,0){$\pg$} 
       \uput{25pt}[180](0.5,1){$m$}
      \psplot{0.505}{0.8}{1 x sub 2 mul ln -1 mul log -1 mul} 
       \listplot[plotstyle=dots]{\circuitmdata}
   \end{pspicture}
   \end{indented}
\caption{Slope $m$ of the logarithmic scaling of saturation depth vs.\
  the fraction $\pg$ of OR gates for $k=2$.  The solid line is the prediction of Eq.~\eqref{eq:pgh}, $m=1/\ln(2(1-\pg))$.}
\label{fig:circuit-m}      
\end{figure}
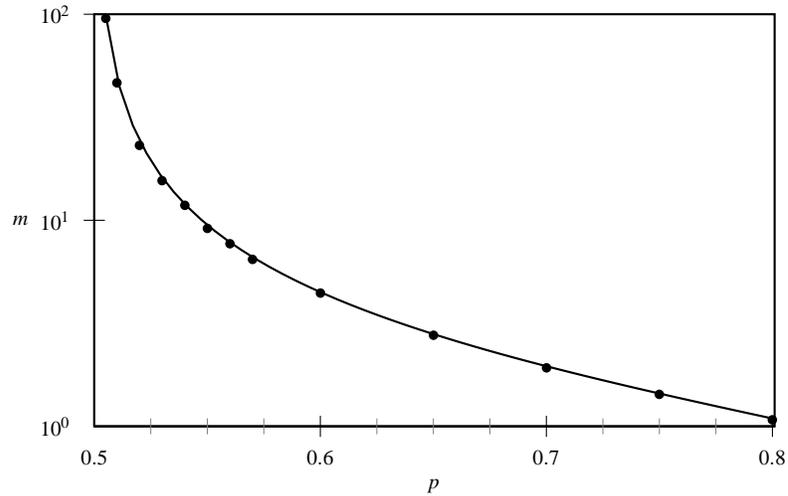

We carried out numerical simulations of random monotone CVP to check the results obtained in Sec.~\ref{sec:2in}.  In our simulations, $\LL$ ranges from 128 to 16384, and we take 1000 realizations of the circuit per data point.  Recall that $\pg$ is the fraction of OR gates.  We first consider two-input gates with $\pg \neq 1/2$.  Figure~\ref{fig:circuit} shows the saturation depth $\T$ versus $\log \LL$ for several values of $\pg$. The initial fraction of \TRUE\ inputs is $\p_0=1/2$. The results demonstrate that the saturation depth increases logarithmically with the circuit width with a slope that increases as $\pg$ approaches the critical value $1/2$.  
Figure~\ref{fig:circuit-m} shows the slope $m = \T / \log \LL$, revealing the divergence as $\pg \rightarrow 1/2$.  The data fits the prediction of Eq.~\eqref{eq:pgh} very well.

\begin{figure}
\begin{indented}
     \item[]
     \psset{xunit=1.8348\linewidth,yunit=0.26\linewidth}
     \begin{pspicture}(0.455,-0.3)(1.01,2)
       \psaxes[xsubticks=4,Dy=1,Dx=0.1,axesstyle=frame,showorigin=true,Ox=0.5]{->}(0.5,0)(1.001,2)
       \uput{20pt}[-90](0.75,0){$\p_0$} 
       \uput{22pt}[180](0.5,1){$m$}
       \psplot{0.501}{0.999}{x ln x mul 1 x sub ln 1 x sub mul add -2 mul} 
       \listplot[plotstyle=dots]{\critmdata}
   \end{pspicture}
   \end{indented}
\caption{Slope $m$ of linear scaling of saturation depth vs.\  the fraction of \TRUE\ inputs $\p_0$ for the critical case, $\pg=1/2$. The solid line is the prediction of Eq.~\eqref{eq:peh}, $m = -2 h(\p_0)$.}                   
\label{fig:crit-m}      
\end{figure}
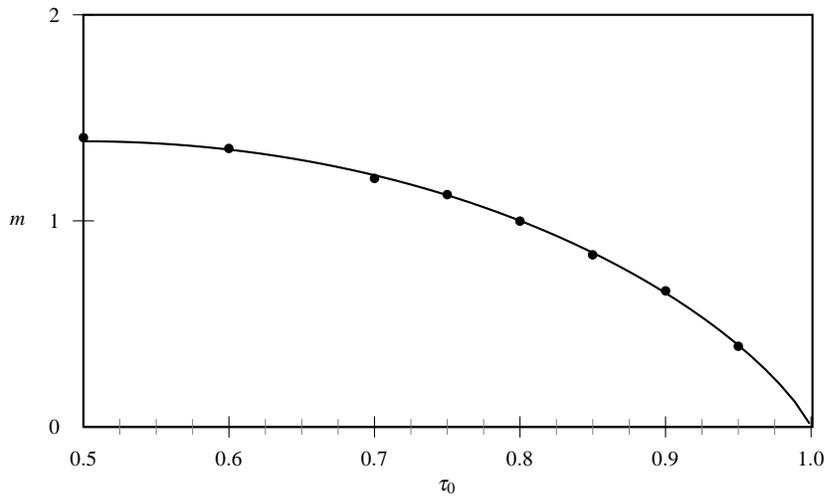

Next we consider the critical case $\pg=1/2$.  Figure~\ref{fig:crit} shows the saturation depth $\T$ as function of $\LL$ for various initial fractions of \TRUE\ inputs $\p_0$.  Note that $\T$ increases linearly with $\LL$ as predicted in Eq.~\eqref{eq:peh}.  The slope $m = \T/\LL$  is shown in Fig.~\ref{fig:crit-m} along with the prediction of Eq.~\eqref{eq:peh}.  The data confirms our estimate based on the mean first-passage time calculation.

\begin{figure}
\begin{indented}
     \item[]
     \psset{xunit=0.000425\linewidth,yunit=0.00019\linewidth}
     \begin{pspicture}(-300,-375)(2000,3000)
      \psaxes[xsubticks=2,Dy=500,Dx=500,axesstyle=frame,showorigin=true]{->}(2000,3000)
       \uput{20pt}[-90](1000,0){$\LL$} 
       \uput{30pt}[180](0,1500){$\T$}
      \psset{showpoints=true}
      \listplot[linecolor=black]{\critAdata}
      \listplot[linecolor=black!90]{\critBdata}
      \listplot[linecolor=black!75]{\critCdata}
      \listplot[linecolor=black!60]{\critDdata}
      \listplot[linecolor=black!45]{\critEdata}
      \listplot[linecolor=black!30]{\critGdata}
     \rput(500,2200){$\p_0 = {\color{black!30}0.5}\ldots 0.95$}
   \end{pspicture}
   \end{indented}
\caption{Saturation depth $\T$ vs. circuit width $\LL$ for various
  values of the fraction of \TRUE\ inputs $\p_0$ for the critical
  case $\pg=0.5$ ($k=2$).}
\label{fig:crit}      
\end{figure}
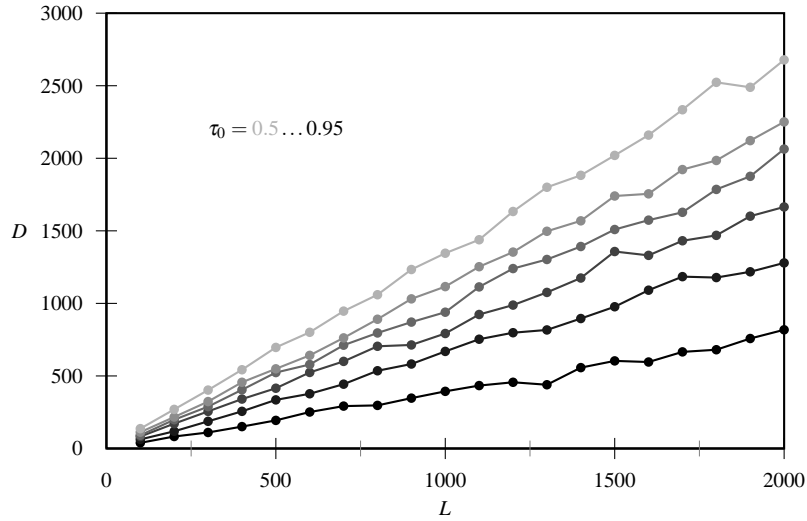

\section{Fast Sampling of Evaluated Random Circuits}
\label{sec:sampling}

In the previous sections we showed that random instances of CVP are hard or easy to solve in parallel depending on the types of connectives and the fraction of connectives of each type.  In this section we consider the complexity of \emph{simultaneously} generating a random instance of CVP, together with its solution---in other words, of sampling the distribution of (instance, solution) pairs.  We show the surprising result that, for any choice of parameters and connectives, random instances of CVP and their solutions can be generated in polylogarithmic parallel time.  The construction depends on generating individual levels of the circuit independently in parallel and then connecting these levels together into a circuit and its solution.  Each level is defined by the placement of each type of connective, the number of \TRUE\ inputs to the level and the evaluation of each connective.   

Here we sketch a polylogarithmic time PRAM program that carries out the construction of a single instance of random CVP together with its solution.  The first step is to generate the inputs to the circuit, $X^i_0 \in \{0,1\}$, $i=1,\ldots,\LL$ and the type of each connectives on each  level $n$, $Y^i_n$ for  $i=1,\ldots,\LL$ where, for example, $Y$ might take the value `three-input OR' or `five-input NAND.'   

The next step in the construction is to evaluate each connective.  Since we don't yet know the number of \TRUE\ inputs $\true_n$ for level $n+1$, we must generate the outputs of all connectives of level $n+1$ for \emph{every} possible number of \TRUE\ inputs that might come from level $n$.   For each $\true_n=0,1,\ldots \LL$, we choose a set of truth values for the inputs of each connective with the correct probability, setting each one \TRUE\ or \FALSE\ with probability $\true_n/L$ or $1-\true_n/L$ respectively.  This determines, for each possible $\true_n$, the outputs of the connectives at level $n+1$.  However, we have not yet chosen the wiring by which these connectives' inputs correspond to connectives on the level $n$.  The construction thus far yields a proto-circuit such as the one shown in Fig.~\ref{fig:all}.

It is important to note that evaluating all possible inputs for a given level does not lead to a combinatorial explosion, since there are only $\LL+1$ possible values for $\true_n$.  Thus we choose just $\LL+1$ instances, or ``possible worlds,'' at each level, and we can do this in parallel with $\bigo(\LL)$ processors.

\newcommand{\gatter}[8]{
   \begin{pspicture}(0.4,0.4)(4.4,3.6)
      \psset{shadow=true,blur=true,shadowsize=2.5pt,blurradius=1.8pt}
      \pnode(0.7,2.8){al} \pnode(1.3,2.8){ar} \pnode(1,1.2){ad}  \cnodeput{0}(1,2){a}{#2} 
      \pnode(2.1,2.8){bl} \pnode(2.7,2.8){br} \pnode(2.4,1.2){bd} \cnodeput{0}(2.4,2){b}{#3} 
      \pnode(3.5,2.8){cl} \pnode(4.1,2.8){cr} \pnode(3.8,1.2){cd} \cnodeput{0}(3.8,2){c}{#4}
      \psset{shadow=false}
      \psframe[framearc=0.2,linewidth=#8](0.4,0.4)(4.4,3.6)
      \ncline{al}{a} \ncline{ar}{a} \ncline{a}{ad}
      \ncline{bl}{b} \ncline{br}{b} \ncline{b}{bd}
      \ncline{cl}{c} \ncline{cr}{c} \ncline{c}{cd}  
      \rput(2.4,3.2){{\tiny #1}}    
      \rput(1,0.8){#5} \rput(2.4,0.8){#6} \rput(3.8,0.8){#7} 
   \end{pspicture}
}
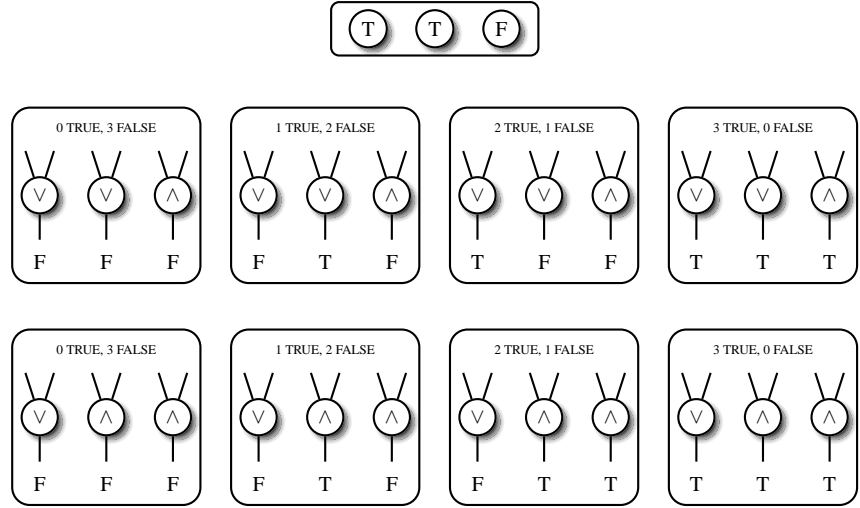
\begin{figure}
\begin{indented}
     \item[]
     \psset{xunit=0.06\linewidth,yunit=0.07\linewidth}
     \begin{pspicture}(0,0)(22.2,10)
       \cnodeput[shadow=true,blur=true,shadowsize=2.5pt,blurradius=1.8pt](6.5,9){ia}{T}
       \cnodeput[shadow=true,blur=true,shadowsize=2.5pt,blurradius=1.8pt](7.9,9){ib}{T}
       \cnodeput[shadow=true,blur=true,shadowsize=2.5pt,blurradius=1.8pt](9.3,9){ic}{F}
       \psframe[framearc=0.3](5.7,8.5)(10.1,9.5)

       \rput(1,6){\gatter{0 \TRUE, 3 \FALSE}{$\vee$}{$\vee$}{$\wedge$}{F}{F}{F}{0.8pt}} 
       \rput(5.6,6){\gatter{1 \TRUE, 2 \FALSE}{$\vee$}{$\vee$}{$\wedge$}{F}{T}{F}{0.8pt}} 
       \rput(10.2,6){\gatter{2 \TRUE, 1 \FALSE}{$\vee$}{$\vee$}{$\wedge$}{T}{F}{F}{0.8pt}}
       \rput(14.8,6){\gatter{3 \TRUE, 0 \FALSE}{$\vee$}{$\vee$}{$\wedge$}{T}{T}{T}{0.8pt}}

       \rput(1,2){\gatter{0 \TRUE, 3 \FALSE}{$\vee$}{$\wedge$}{$\wedge$}{F}{F}{F}{0.8pt}} 
       \rput(5.6,2){\gatter{1 \TRUE, 2 \FALSE}{$\vee$}{$\wedge$}{$\wedge$}{F}{T}{F}{0.8pt}} 
       \rput(10.2,2){\gatter{2 \TRUE, 1 \FALSE}{$\vee$}{$\wedge$}{$\wedge$}{F}{T}{T}{0.8pt}}
       \rput(14.8,2){\gatter{3 \TRUE, 0 \FALSE}{$\vee$}{$\wedge$}{$\wedge$}{T}{T}{T}{0.8pt}}
     \end{pspicture}
   \end{indented}
\caption{A proto-circuit composed of 2-input AND and OR gates of width $\LL=3$ and two levels.   For each level and each number of \TRUE\ inputs to a level, each gate in the level is independently evaluated.}                   
\label{fig:all}      
\end{figure}

\begin{figure}
\begin{indented}
     \item[]
     \psset{xunit=0.06\linewidth,yunit=0.07\linewidth}
     \begin{pspicture}(0,0)(22.2,10)
       \cnodeput[shadow=true,blur=true,shadowsize=2.5pt,blurradius=1.8pt](6.5,9){ia}{T}
       \cnodeput[shadow=true,blur=true,shadowsize=2.5pt,blurradius=1.8pt](7.9,9){ib}{T}
       \cnodeput[shadow=true,blur=true,shadowsize=2.5pt,blurradius=1.8pt](9.3,9){ic}{F}
       \psframe[framearc=0.3,linewidth=2pt](5.7,8.5)(10.1,9.5)
       \psline[linewidth=2pt,ArrowInside=->,arrowsize=0.3]{-}(7.9,8.5)(10.2,7.6)
       \psline[linewidth=2pt,ArrowInside=->,arrowsize=0.3]{-}(10.2,4.4)(5.6,3.6)

       \rput(1,6){\gatter{0 \TRUE, 3 \FALSE}{$\vee$}{$\vee$}{$\wedge$}{F}{F}{F}{0.8pt}} 
       \rput(5.6,6){\gatter{1 \TRUE, 2 \FALSE}{$\vee$}{$\vee$}{$\wedge$}{F}{T}{F}{0.8pt}} 
       \rput(10.2,6){\gatter{2 \TRUE, 1 \FALSE}{$\vee$}{$\vee$}{$\wedge$}{T}{F}{F}{2pt}}
       \rput(14.8,6){\gatter{3 \TRUE, 0 \FALSE}{$\vee$}{$\vee$}{$\wedge$}{T}{T}{T}{0.8pt}}

       \rput(1,2){\gatter{0 \TRUE, 3 \FALSE}{$\vee$}{$\wedge$}{$\wedge$}{F}{F}{F}{0.8pt}} 
       \rput(5.6,2){\gatter{1 \TRUE, 2 \FALSE}{$\vee$}{$\wedge$}{$\wedge$}{F}{T}{F}{2pt}} 
       \rput(10.2,2){\gatter{2 \TRUE, 1 \FALSE}{$\vee$}{$\wedge$}{$\wedge$}{F}{T}{T}{0.8pt}}
       \rput(14.8,2){\gatter{3 \TRUE, 0 \FALSE}{$\vee$}{$\wedge$}{$\wedge$}{T}{T}{T}{0.8pt}}
     \end{pspicture}
   \end{indented}
\caption{Pruning the proto-circuit.  Instances of each level are consistently connected to the succeeding level and the subgraph connected to the circuit input is identified.}                   
\label{fig:prune}      
\end{figure}
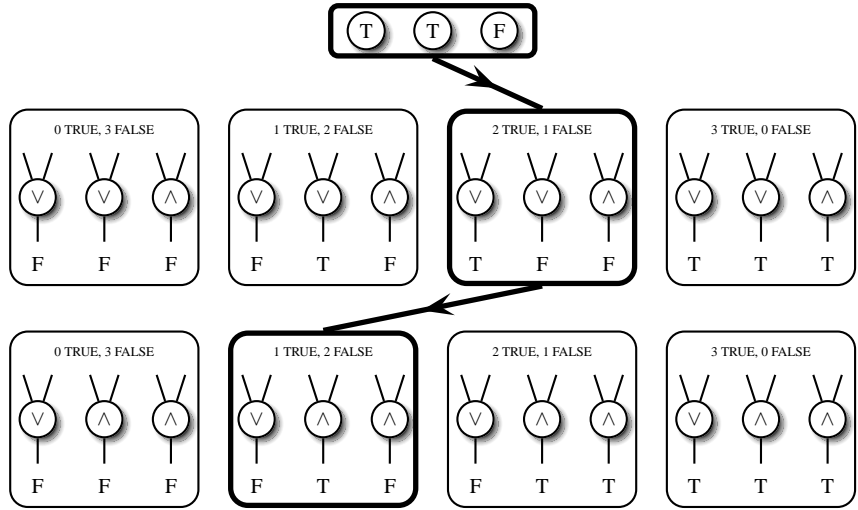

The next step in the construction is to connect these possible worlds in a consistent way.  Figure~\ref{fig:prune} shows how to do this.   We create a directed graph where each vertex $(n,m)$ corresponds to the instance of level $n$ with $m$ \TRUE\ inputs.  Having chosen what the outputs of the connectives will be for each value of the inputs, we draw an edge from $(n,m)$ to the corresponding vertex $(n+1,\true_{(n,m)})$ where $\true_{(n,m)}$ is the number of \TRUE\ outputs of the instance of level $n$ with $m$ \TRUE\ inputs.  
A logically consistent history is then the unique directed path starting at $(0,\true_0)$, 
shown in Fig.~\ref{fig:prune}.  We can find paths through directed graphs in polylogarithmic parallel time as a function of the total number $N$ of vertices~\cite{GiRy}.  If the circuit has width $\LL$ and depth $n$, then $N=n(\LL+1)$.

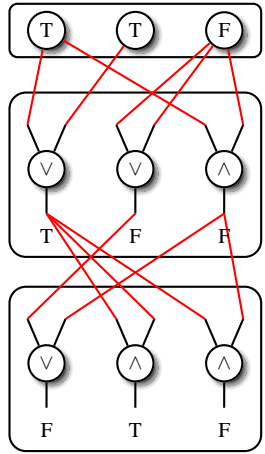
\begin{figure}
\begin{indented}
     \item[]
     \begin{center}
     \psset{xunit=0.08\linewidth,yunit=0.07\linewidth}
     \begin{pspicture}(0,0)(5,9)
       \cnodeput[shadow=true,blur=true,shadowsize=2.5pt,blurradius=1.8pt](1.1,8){ia}{T}
       \cnodeput[shadow=true,blur=true,shadowsize=2.5pt,blurradius=1.8pt](2.5,8){ib}{T}
       \cnodeput[shadow=true,blur=true,shadowsize=2.5pt,blurradius=1.8pt](3.9,8){ic}{F}
       \psframe[framearc=0.3](0.5,7.5)(4.5,8.5)
      
       \psset{shadow=true,blur=true,shadowsize=2.5pt,blurradius=1.8pt}
       \pnode(0.8,2.8){alo} \pnode(1.4,2.8){aro} \pnode(1.1,1.2){ado}  \cnodeput{0}(1.1,2){ao}{$\vee$} 
       \pnode(2.2,2.8){blo} \pnode(2.8,2.8){bro} \pnode(2.5,1.2){bdo} \cnodeput{0}(2.5,2){bo}{$\wedge$} 
       \pnode(3.6,2.8){clo} \pnode(4.2,2.8){cro} \pnode(3.9,1.2){cdo} \cnodeput{0}(3.9,2){co}{$\wedge$}
       \psset{shadow=false}
       \psframe[framearc=0.2,linewidth=0.8pt](0.5,0.4)(4.5,3.4)
       \ncline{alo}{ao} \ncline{aro}{ao} \ncline{ao}{ado}
       \ncline{blo}{bo} \ncline{bro}{bo} \ncline{bo}{bdo}
       \ncline{clo}{co} \ncline{cro}{co} \ncline{co}{cdo}  
       \rput(1.1,0.8){F} \rput(2.5,0.8){T} \rput(3.9,0.8){F}

       \psset{shadow=true,blur=true,shadowsize=2.5pt,blurradius=1.8pt}
       \pnode(0.8,6.3){al} \pnode(1.4,6.3){ar} \pnode(1.1,4.7){ad}  \cnodeput{0}(1.1,5.5){a}{$\vee$} 
       \pnode(2.2,6.3){bl} \pnode(2.8,6.3){br} \pnode(2.5,4.7){bd} \cnodeput{0}(2.5,5.5){b}{$\vee$} 
       \pnode(3.6,6.3){cl} \pnode(4.2,6.3){cr} \pnode(3.9,4.7){cd} \cnodeput{0}(3.9,5.5){c}{$\wedge$}
       \psset{shadow=false}
       \psframe[framearc=0.2,linewidth=0.8pt](0.5,3.9)(4.5,6.9)
       \ncline{al}{a} \ncline{ar}{a} \ncline{a}{ad}
       \ncline{bl}{b} \ncline{br}{b} \ncline{b}{bd}
       \ncline{cl}{c} \ncline{cr}{c} \ncline{c}{cd}  

       \psset{linecolor=red}
       \ncline{ia}{al} \ncline{ia}{cl} \ncline{ib}{ar} 
       \ncline{ic}{bl} \ncline{ic}{br} \ncline{ic}{cr}
       \ncline{ad}{blo} \ncline{ad}{bro} \ncline{bd}{alo} \ncline{ad}{clo}
       \ncline{cd}{aro} \ncline{cd}{cro}

       \psset{linecolor=black}
       \rput(1.1,4.3){T} \rput(2.5,4.3){F} \rput(3.9,4.3){F}
    \end{pspicture}
    \end{center}
   \end{indented}
\caption{The circuit and its solutions is generated by randomly connecting gates at successive levels consistent with the given output of each gate.}                   
\label{fig:final}      
\end{figure}

The final step in constructing the circuit is randomly connecting each connective to the ones in the previous level.  Having chosen the truth values of its inputs, we simply choose each of its \TRUE\ inputs randomly with replacement from the connectives with \TRUE\ outputs at the previous level, and similarly for its \FALSE\ inputs.  This can be carried out independently in parallel for each connective.  The result of this procedure is a properly sampled circuit and its solution, i.e., a pair (instance, solution) of random CVP, as shown in Fig.~\ref{fig:final}.  The entire construction requires polylogarithmic parallel time on a PRAM with polynomially many processors.  

We emphasize that we can carry out this construction even if the types of the connectives are not independent and identically distributed at each level.  This distribution can vary from level to level, or even be highly correlated within or between levels, as long as the joint distribution of the set of $nL$ connective types can be sampled in \NC: that is, as long as there is a PRAM program that runs in $\bigo(\log^k n\LL)$ time, for some constant $k$, that takes a seed with $\poly(n\LL)$ random bits and produces a sample from the joint distribution.  The only randomness we need for the construction to work is that the inputs to each connective are chosen uniformly and with replacement from those at the previous level.

\section{Discussion}
\label{sec:disc}

We have studied several random ensembles of feedforward Boolean circuits and found that, depending on the types of connectives and the fraction of connectives of each type, the circuit may be easy or hard to evaluate in parallel.  The easy circuits rapidly saturate to a single truth value or, in the case of NOR circuits, a period-two oscillation between \TRUE\ and \FALSE. For these ensembles, it is only necessary to evaluate the circuit to logarithmic depth to learn its ultimate output, since nothing changes after saturation has occurred.  On the other hand, for other choices of random ensembles, saturation occurs slowly, if at all, and circuit evaluation is presumably hard to carry out in parallel.  Thus, although the monotone and NOR Circuit Value Problems studied here are all \PP-complete, this worst-case classification does not distinguish among different possibilities for the average case complexity of parallel evaluation of the circuit.

 When there is a single attracting fixed point, as was found for monotone circuits with more than two inputs, the exact evaluation of the circuit is to hard to accomplish in parallel but the statistical properties of the outputs at each level are predictable and insensitive to the specific inputs and wiring of the circuit.  For more complicated connectives, including some with just five inputs, the recursion relations for the expected number of \TRUE\ values on a level can lead to chaotic dynamics.  For these ensembles of random circuits, even the statistical properties of the outputs of each level are hard to predict since the fraction of \TRUE\ outputs at each level is extremely sensitive to the initial truth values and the wiring of the circuit. 

In contrast to evaluating a given instance of a Boolean circuit, which may be easy or hard to accomplish in parallel, it is \emph{always} easy  to sample in parallel evaluated instances chosen from a random ensemble---that is, (instance, solution) pairs.  The underlying idea is that the output of each connective can be chosen independently from the correct probability distribution as a function of the number of \TRUE\ inputs in the previous level.  Once the full set of possibilities is generated for each level, a consistent history is a path through a directed graph of polynomial size, and the whole construction of the circuit can be completed in polylogarithmic parallel time on a PRAM.  One interesting consequence of this result, combined with the existence of chaotic connectives, is that it is possible to generate a chaotic sequence of numbers in a time that is polylogarithmic in the length of sequence.  

In statistical physics, we are often concerned with sampling ensembles of random instances of problems together with their solutions.  The sampling result for random Boolean circuits holds out the promise that for other problems of interest in statistical physics, it may be possible to sample instances together with solutions with less computational effort than the traditional method of first generating an instance and then solving it.  

\ack
J.\ M.\ was supported in part by NSF Grant DMR-0907235.  C.\ M. was supported by the McDonnell Foundation.  S.\ M. was supported by the European CommunityÕs FP6 Information Society Technologies program, contract IST-001935, EVERGROW.

\section*{References}

\bibliographystyle{unsrt} 
\bibliography{rancvp}

\end{document}